\documentclass[sigconf]{acmart}

\AtBeginDocument{%
  \providecommand\BibTeX{{%
    \normalfont B\kern-0.5em{\scshape i\kern-0.25em b}\kern-0.8em\TeX}}}

\copyrightyear{2022} 
\acmYear{2022} 
\setcopyright{acmcopyright}
\acmConference[WWW '22]{Proceedings of the ACM Web Conference 2022}{April 25--29, 2022}{Virtual Event, Lyon, France}
\acmBooktitle{Proceedings of the ACM Web Conference 2022 (WWW '22), April 25--29, 2022, Virtual Event, Lyon, France}
\acmPrice{15.00}
\acmDOI{10.1145/3485447.3511977}
\acmISBN{978-1-4503-9096-5/22/04}

\usepackage{graphicx}
\usepackage{multirow}
\usepackage{pgfplots}
\usepackage{subfigure}
\usepackage{caption}
\pgfplotsset{compat=1.16}

\begin{document}

\title{A Multi-task Learning Framework for Product Ranking with BERT}

\author{Xuyang Wu}
\affiliation{%
  \institution{Santa Clara University}
  \city{Santa Clara}
  \state{CA}
  \country{USA}
}
\email{xwu5@scu.edu}

\author{Alessandro Magnani}
\affiliation{%
  \institution{Walmart Global Tech}
  \city{Sunnyvale}
  \state{CA}
  \country{USA}}
\email{alessandro.magnani@walmart.com}

\author{Suthee Chaidaroon}
\affiliation{%
  \institution{Walmart Global Tech}
  \city{Sunnyvale}
  \state{CA}
  \country{USA}
}
\email{suthee.chaidaroon@walmart.com}

\author{Ajit Puthenputhussery}
\affiliation{%
 \institution{Walmart Global Tech}
  \city{Sunnyvale}
  \state{CA}
 \country{USA}}
 \email{ajit.puthenputhussery@walmart.com}

\author{Ciya Liao}
\affiliation{%
  \institution{Walmart Global Tech}
  \city{Sunnyvale}
  \state{CA}
  \country{USA}}
 \email{ciya.liao@walmart.com} 

\author{Yi Fang}
\affiliation{%
  \institution{Santa Clara University}
  \city{Santa Clara}
  \state{CA}
  \country{USA}}
\email{yfang@scu.edu}

\begin{abstract}
Product ranking is a crucial component for many e-commerce services. One of the major challenges in product search is the vocabulary mismatch between query and products, which may be a larger vocabulary gap problem compared to other information retrieval domains. While there is a growing collection of neural learning to match methods aimed specifically at overcoming this issue, they do not leverage the recent advances of large language models for product search. On the other hand, product ranking often deals with multiple types of engagement signals such as clicks, add-to-cart, and purchases, while most of the existing works are focused on optimizing one single metric such as click-through rate, which may suffer from data sparsity. In this work, we propose a novel end-to-end multi-task learning framework for product ranking with BERT to address the above challenges. The proposed model utilizes domain-specific BERT with fine-tuning to bridge the vocabulary gap and employs multi-task learning to optimize multiple objectives simultaneously, which yields a general end-to-end learning framework for product search. We conduct a set of comprehensive experiments on a real-world e-commerce dataset and demonstrate significant improvement of the proposed approach over the state-of-the-art baseline methods.

\end{abstract}



\maketitle

\section{Introduction}

Online shopping has now become an integral part of people’s daily life. With an ever increasing catalog size, product search systems have been playing a crucial role in serving customers shopping on online e-commerce platforms, by benefiting both users and suppliers. Many approaches have been proposed for product search, ranging from adaptations of general web search models \cite{b45} to applying learning to rank to the e-commerce domain \cite{b36}. Recently, neural information retrieval (IR) methods based on deep neural networks have been introduced and utilized in product search \cite{b40,b44}, which has demonstrated promising results by learning semantic representations of queries and products. There exist several challenges in applying neural IR to product search. First of all, product search often deals with multiple types of engagement signals such as clicks, add-to-cart, and purchases, while most of the existing work is focused on optimizing one single metric such as click-through rate. In addition, neural IR generally requires a huge amount of training data and one source of data may not be able to support sufficient model training. Last but not least, the vocabulary mismatch between query and document in product search may be a larger problem than in other IR domains \cite{b47}. The existing work does not fully utilize the recent advances of large language models for product search as mediocre results are reported when using BERT in product ranking \cite{b35}.

In this paper, we tackle the above challenges by proposing a multi-task learning framework for product ranking with BERT. The proposed framework aims at optimizing multiple objectives at the same time such as click, add-to-cart, and purchase. One of the major challenges in applying multi-task learning is that the intrinsic conflicts between different tasks can hurt the model performance on some of the tasks, especially when model parameters are heavily shared among the tasks. As a result, many deep learning based multi-task learning models are sensitive to factors such as the data distributions and relationships among tasks \cite{b4}. Our proposed multi-task learning architecture was inspired by the Mixture-of-Experts (MoE) layer \cite{b48} and the recent MMoE model \cite{b4}. Compared with roughly shared parameters in MMoE, we explicitly separate shared and task-specific experts to alleviate harmful parameter interference between common and task-specific knowledge. 

Given a user query, a ranked list of products is shown and the user may click on the relevant items, add some of them to the shopping cart, and eventually purchase one or more items. 
The user engagement action generally follows a sequential order, impression $\rightarrow$ click $\rightarrow$ add-to-cart $\rightarrow$ purchase.
In other words, the later step may occur only if the former steps are completed with positive feedback. 
In this paper, we utilize probability transfer \cite{b6} over the sequential user behaviors to leverage all impression samples over the entire space and extra abundant supervisory signals from post-click behaviors, efficiently addressing the data sparsity issue. Specifically, we model the respective probabilities of click, add-to-cart, and purchase simultaneously in a multi-task learning framework according to the conditional probability rule defined on the user behavior graph.

On the other hand, the vocabulary mismatch between query and document poses a fundamental challenge in search engines in general. In product search, the vocabulary gap may be a larger problem than in other information retrieval domains \cite{b35,b47}. Queries and product titles are usually short, and titles are sometimes phrases or simple combinations of keywords instead of well-structured sentences. There is a growing interest in utilizing neural learning to match methods for bridging the vocabulary gap in product search. These methods go beyond lexical matching by representing queries and documents as dense semantic vectors and learning their degree of similarity in the semantic space \cite{b42}. However, the existing work has not fully utilized the advantages of large pre-trained language models such as BERT \cite{b13}.
A recent study has explored the state-of-the-art BERT-based model for product search but demonstrated mediocre results when comparing with traditional methods \cite{b35}. 
In this paper, we pre-train a domain-specific BERT model with fine-tuning and show much promising results. The main contributions of the paper can be summarized as follow:

\begin{itemize}
\item We propose a novel multi-task learning framework for product ranking based on neural information retrieval. To the best of our knowledge, there exists no prior work on integrating multiple types of engagement signals with neural IR for product search.
\item To model the relationships between different tasks and alleviate harmful parameter interference between common and task-specific knowledge, the proposed framework utilizes the mixture-of-experts layer and probability transfer between tasks to harness the abundant engagement data.
\item The proposed framework integrates semantic match with traditional ranking features in an end-to-end learning manner. We leverage a domain-specific BERT for semantic match with fine-tuning and demonstrate promising results. 

\item We conduct a comprehensive set of experiments on a real-world product search dataset in e-commerce and show the effectiveness of the proposed approach over the competitive baseline methods. 

\end{itemize}

\section{Related Work}

\subsection{Product Search and Ranking}

Product search problem is more challenging than the traditional web search problem \cite{b45} as the user queries tend to be short and there are millions of potentially related products \cite{b35}. Some researchers proposed an iterative approach with two or more steps where an initial set of candidate items is retrieved. The retrieved items were then iteratively ordered (re-ranked) and reduced in size by picking the top ones in the list \cite{b36}. Long et al. \cite{b37} proposed a model that incorporates best-selling products data to rank the search results. Also, the model in \cite{b38} considered the diversity of the product results to enhance the user experience. As various signals could be used in e-commerce to measure the quality of the search results, some works \cite{b39, b41} optimized the search results based on user engagements, such as click-through rate and conversion rate. Unfortunately, user engagement data, is often very sparse and it can therefore limit the performance of the model. Deep neural-based models \cite{b42, b43} were also used in both the retrieval and ranking steps. Magnani et al. \cite{b40} enhanced the deep learning-based model using different types of text representation and loss function. Zhang et al. \cite{b44} added to the ranking model, interaction features between the query and a graph of products that captured relationship between similar products. 

\subsection{Neural Information Retrieval}

In neural information retrieval, ranking models can be categorized into two main groups, representation-based and interaction-based models. Representation-based models learn an embedding for queries and items respectively and then measure the relevance of a product for a given query, by computing a distance between the query and item embedding. DSSM \cite{b12} computed query and item embeddings by averaging word embeddings from query text and document text fields, respectively. CLSM \cite{b30} and LSTM-RNN \cite{b22} used CNN \cite{b32} and LSTM \cite{b33} networks. NRM-F \cite{b34} achieved a better performance by encoding multiple product fields (e.g. title, description, color) in the embedding. Representation-based methods were often limited by the embedding size and consequently they could not summarize all the information from the original data and capture precise lexical matching.

Interaction-based models looked at interactions between word pairs in queries and items. DRMM \cite{b19} computed the cosine similarity between each word embedding in the query and item. In recent years,  BERT-based \cite{b13} models have achieved the state-of-the-art performance \cite{b14,b15,b16} in ranking. They usually concatenated the query and document string as one sentence which was then fed to multiple transformer layers. The attention mechanism \cite{b17} within the network could attend to each word pair of the query and product. 

\subsection{Multi-task Learning}

Multi-Task learning aims to improve generalization by leveraging domain-specific information in the training signals of related tasks \cite{b26}. It has several advantages over traditional single-task learning. Due to their inherent layer sharing, the resulting memory efficiency can be substantially reduced and the inference speed can be improved. Moreover, the associated tasks can benefit from each other if they share complementary information, or act as a regularizer for one another.

The early multi-task learning (MTL) work mainly focused on hard parameter sharing \cite{b27}. This is also a very common type of MTL models. The output of the shared layers fed unique modules for different tasks. When tasks were highly correlated, this structure could often achieve good results. When the tasks were not so correlated, there could be a negative migration phenomenon. Some works, such as MMoE \cite{b4} and PLE \cite{b5}, addressed this issue by utilizing multiple experts on a shared bottom structure. Based on the gating mechanism, different tasks could filter the output of different experts, shared experts, and task-specific experts. This type of models mainly learned in the shared architecture at the bottom but did not exchange more information at the top.

Some other ideas, such as ESMM \cite{b2} and $ESM^2$ \cite{b6} models, used probability transfer based on the sequential order between different tasks at the top of the model to optimize the model effect and achieve better results in click-through rate and conversion rate estimation tasks. In \cite{b23, b28}, neural collaborative filtering was extended to the setting of multi-task learning. The above models mainly used probability transfer, in which only simple probability information was transferred between adjacent tasks. Xi et al. \cite{b1} proposed AITM, which modeled the sequential dependence among multi-step conversions and adaptively learned what and how much information to transfer for different conversion stages.

\section{Multi-task Learning For Product Ranking}

\begin{figure*}[htbp]
\centering
\includegraphics[width=0.90\textwidth]{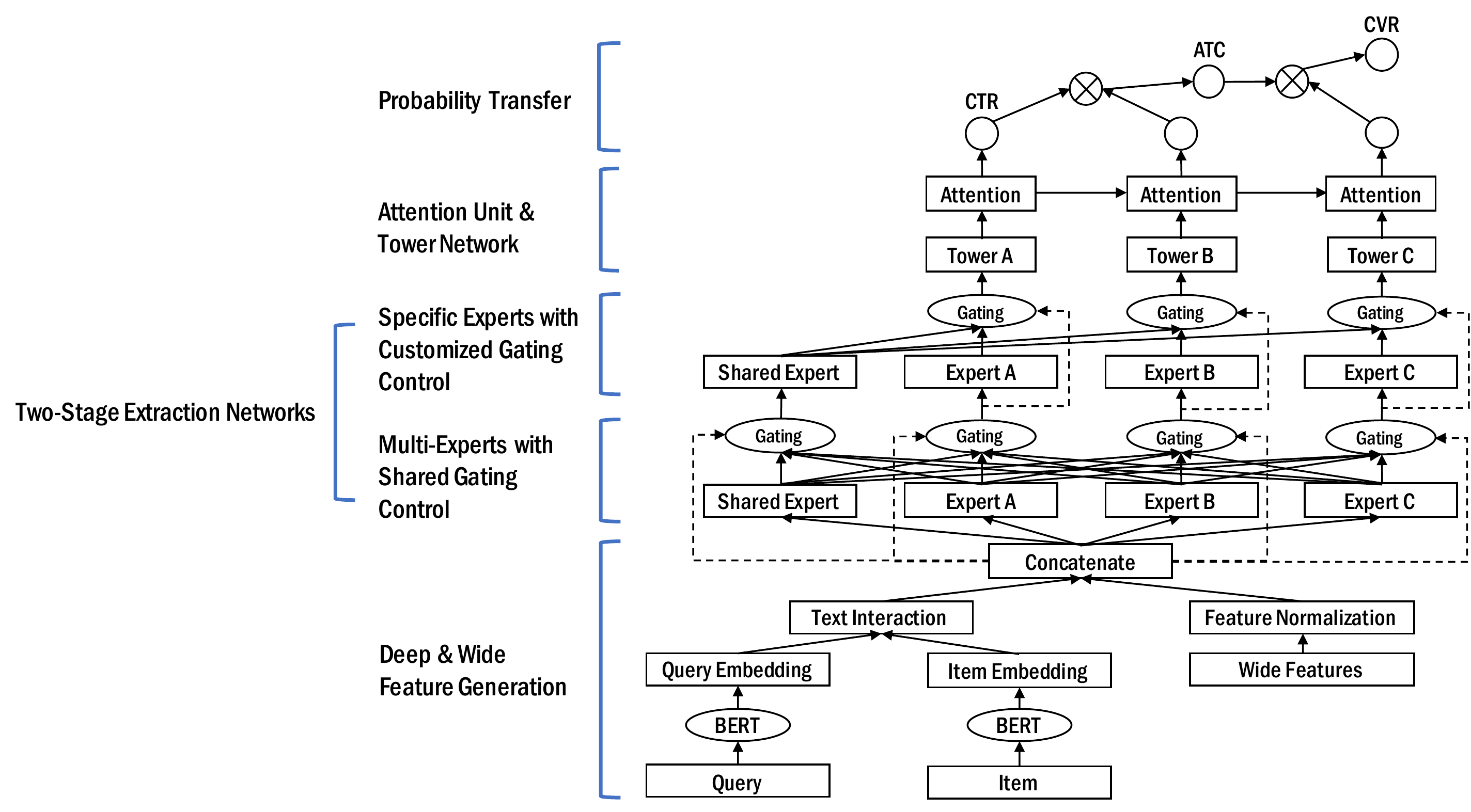}
\caption{The architecture of the proposed multi-task learning framework for product ranking (MLPR).}
\label{Fig.framework} 
\end{figure*}

In this section, we present the proposed multi-task learning framework for product ranking (MLPR). Let $\mathcal{Q}=\{q_1, q_2,...,q_n,..., q_N\}$ denote the collection of $N$ user queries, and $\mathcal{I}=\{i_1, i_2,...,i_m,..., i_M\}$ denote the collection of $M$ products (items). 
Given the search results, we consider three types of user activities: click, add-to-cart (ATC), and purchase, which follow a sequential order as follow: \textit{impression} $\rightarrow$ \textit{click} $\rightarrow$ \textit{add-to-cart} $\rightarrow$ \textit{purchase}. 
Our task is to predict the probability of each engagement activity given a query $q_n$ and product $i_m$ pair, mathematically as follow

\begin{equation}
\begin{aligned}
\hat{y}^k_{n,m}=\mathcal{F}^k(\phi(q_n), \psi(i_m))
\end{aligned}
\end{equation}
where $\phi(\cdot)$ and $\psi(\cdot)$ denote the query encoder and the item encoder, respectively. $\mathcal{F}^k(\cdot)$ denotes the prediction probability function for the task $k$ based on query $q_n$ and product $i_m$ pair. Since there are three types of engagement activities, we formulate them as multi-task learning problem by optimizing all these three objectives simultaneously.

Figure \ref{Fig.framework} illustrates the architecture of the proposed framework, which consists of five stages: deep \& wide feature generation, multi-experts with shared gating control, specific-experts with customized gating control, tower network and attention unit, and probability transfer. Given $K$ tasks, the deep \& wide feature generation stage create the input features based on the raw data, which is followed by the two-stage extraction networks designed as a shared-bottom structure. The tower networks with the attention units are built upon the output of the  extraction networks. They generate the output for the corresponding task $k$. We will present each stage in details in the following subsections.

\subsection{Deep \& Wide Feature Generation}

The deep features include query embedding, product embedding, and their interactions. In MLPR, we leverage a domain-specific BERT for learning the embeddings, which are pre-trained on the e-commerce domain data. The query embedding is generated by the domain-specific BERT from the query text field. The product embedding is generated from the title field, type field, brand field, color field, and gender field of the product. After obtaining the embedding features, we also compute the interactions between query embedding and product embedding based on Cosine similarity $\dfrac{\mathbf{\mu}_q^T\cdot\mathbf{\mu}_i}{\parallel\mathbf{\mu}_q\parallel\cdot\parallel\mathbf{\mu}_i\parallel}$, Hadamard (element-wise) product $\mathbf{\mu}_q\circ\mathbf{\mu}_i$, and concatenation $\mathbf{\mu}_q\oplus\mathbf{\mu}_i$.

\paragraph{\textbf{Domain-specific BERT}} We utilize the fine-tuned BERT model to generate the input query and item embedding. We first initialized the BERT model with the pre-trained weights taken from the distillbert-base-uncased\footnote{\url{https://huggingface.co/distilbert-base-uncased}}. Then the BERT model was  fine-tuned based on the user engagement logs collected from the e-commerce website. Each row of the log file consists of one query and a list of clicked, add-to-cart and purchased items. The training objective is to estimate the optimal order of these items by using the numbers of clicks, add-to-cart and purchases as the ground truth. For each query, we also injected randomly sampled items. The ratio of relevant and sampled items are 1:20. We use a raw query and item attributes such as title, color, brand, and product types as the inputs to the model. The model has both query and item encoders. The last layer of the encoder outputs 256-dimensional query and item vectors.

Wide features directly come from the production side ranking features, which are the traditional features used in learning to rank. They can be generally grouped into the following categories: query item level engagement features (e.g., query item CTR, ATC, order ratio, etc.), item attributes (e.g., category, price, rating score, review count, etc.), iteration features (e.g., similarity score, matching score) and so on. In our experiments, we remove the engagement ranking features to avoid any potential data leak. We obtained a total of 243 ranking features. In addition, we use the z-score to normalize all the ranking features with mean and standard deviation computed from training data. The concatenated features after deep \& wide feature generation are denoted as follows:

\begin{equation}
    \mathbf{x}_{n,m} = concat(\phi(q_n), \psi(i_m), \varphi(\phi(q_n), \psi(i_m)), \mathbf{r}_{n,m}),
\end{equation}
where $\phi(\cdot)$, $\psi(\cdot)$ and $\varphi(\cdot)$ denote the query encoder, item encoder and interaction feature encoder, respectively. $\mathbf{r}_{n,m}$ denotes the ranking features for each query and item pair $(q_n, i_m)$. $\mathbf{x}_{n,m}$ denote the concatenated feature vector generated from query and item input field. 

\subsection{Two-stage Extraction Networks}

Many multi-task learning models in the existing work contain a shared layer at the bottom, which can learn common knowledge from different tasks. In addition, the shared experts can continuously absorb the joint hidden information from different tasks. This structure may help alleviate overfitting, but it could negatively affect model performance due to task dependence and data distribution. We propose two stages in the the extraction networks and explicitly separate shared
and task-specific experts to avoid harmful parameter interference.

\subsubsection{Multi-Experts with Shared Gating Control}

In this stage, the model utilizes the gating network mechanism at the bottom of the model based on the principle of MMoE \cite{b4}. Each task uses a separate gating network. The gating network of each task achieves the selective utilization in different task networks through different final output weights. Various schemes of gating networks can learn different patterns of combined experts, and thus the model will consider the relevance and difference of each task. For each input from the previous stage, the current stage can select the partial meaningful experts by the gating network conditioned on the input. Each expert network is a simple multi-layer feed-forward network with batch normalization and ReLu activation function. The gating network is designed as a single-layer feed-forward network with a Softmax activation function. The output of each gating network is formulated as:
    
\begin{equation}
    \label{eq:3}
    \begin{aligned}
        w^k(\mathbf{x}_{n,m}) &= Softmax(\mathbf{W}^k_g\mathbf{x}_{n,m}) \\
        g^k(\mathbf{x}_{n,m}) &= w^k(\mathbf{x}_{n,m})E^k(\mathbf{x}_{n,m})
    \end{aligned}
\end{equation}
where $\mathbf{x}_{n,m}$ is the concatenated feature vector from the deep \& wide feature generation layer, $\mathbf{W}^k_g$ is the trainable parameter matrix for task $k$, $w^k(\mathbf{x}_{n,m})$ is the weighting function which obtains the weighted vector of task $k$ by a linear layer with the Softmax activation function. $E^k(\mathbf{x}_{n,m})$ is the expert network.

\subsubsection{Specific-Experts with Customized Gating Control}

In this stage, our model applies the specific experts with customized gating controllers to extract the task-specific hidden information. The shared expert module and the task-specific expert module will obtain the input from the previous stage. The parameters in the shared expert are affected by all the tasks. They are in the task-specific expert affected by the corresponding task \cite{b5}. 

\begin{equation}
    \label{eq:4}
    \begin{aligned}
        w^k(\mathbf{x}) &= Softmax(\mathbf{W}^k_g \mathbf{x}) \\
        H^k(\mathbf{x}) &= {[E^k(\mathbf{x}), E^s(\mathbf{x})]}^T \\
        g^k(\mathbf{x}) &= w^k(\mathbf{x})H^k(\mathbf{x})
    \end{aligned}
\end{equation}
where $H^k(\mathbf{x})$ denotes the vector of the combination of shared experts $E^s(\mathbf{x})$ and the task $k$'s specific experts $E^k(\mathbf{x})$, $\mathbf{x}$ denotes the previous layer's output $g^k(\mathbf{x}_{n,m})$. Then the model uses the gating network to calculate the weighted sum of the selected vectors, which is the same structure as Eqn. (\ref{eq:3}) in the previous stage with different parameter matrix $\mathbf{W}^k_g$ and input experts $H^k(\mathbf{x})$.
    

\subsection{Tower Network \& Attention Unit}
In the upper stage, tower networks obtain the prediction corresponding to each task. Each tower network is a simple multi-layer feed-forward network, and it can be extended to any advanced structure. The attention units learn more task-driven confidential information within the tower network. For a task $k$, those units could adaptively transfer helpful information from the former task. 
Given the $K$ tasks, the output $\mathbf{t}^k$ of the tower network for each task $k$ is defined as follows:
\begin{equation}
\mathbf{t}^k = {MLP}^k(\mathbf{v})
\end{equation}
where $\mathbf{t}^k(\cdot)$ denotes the tower network and input $\mathbf{v}$ is the output of the shared-bottom stage, the output of Eqn. (\ref{eq:4}).

For the attention units, there are two inputs from the adjacent tasks $k-1$ and $k$, respectively, and the output of attention unit $\mathbf{a}^k$ of the task $k$ is defined as:

\begin{equation}
\mathbf{a}^k = Attention(\mathbf{t}^k, \mathbf{a}^{k-1})=softmax(\frac{\mathbf{QK}^T}{\sqrt{d_k}})\mathbf{V}
\label{att}
\end{equation}
where $Attention(\cdot, \cdot)$ function is the similar design with self-attention mechanism \cite{b17}, and $\mathbf{t}^k$ is the tower network's output, $\mathbf{a}^{k-1}$ is the attention unit output from the former task. $\mathbf{Q}=\mathbf{W_Q}(\mathbf{t}^k\oplus\mathbf{a}^{k-1}), \mathbf{K}=\mathbf{W_K}(\mathbf{t}^k\oplus\mathbf{a}^{k-1}), \mathbf{V}=\mathbf{W_V}(\mathbf{t}^k\oplus\mathbf{a}^{k-1})$ is a simple single-layer feed-forward network with different weight matrix $\mathbf{W_Q,W_K,W_V}$, respectively. For the first task without former task, $\mathbf{a}^1 = Attention(\mathbf{t}^1, \emptyset)$.

The output of attention unit $a^k$ feeds into a single-layer feed-forward network ${MLP}^k(\cdot)$ to obtain the corresponding prediction probability $\hat{p}^k$ for each task $k$.
\begin{equation}
    \hat{p}^k = sigmoid({MLP}^k(\mathbf{a}^k))
\end{equation}

\subsection{Probability Transfer}
To alleviate the data sparsity and the bias of sample space, the proposed framework adopts the probability transfer mechanism \cite{b2}, which is defined on the user behavior graph \textit{impression} $\rightarrow$ \textit{click} $\rightarrow$ \textit{add-to-cart} $\rightarrow$ \textit{purchase}. Given the impression $\mathbf{x}$, the model prediction probability transfer is defined as:

\begin{equation}
\begin{aligned}
\hat{y}^{Click} &= \hat{p}^{ctr} = p(y^{Click}=1|\mathbf{x}) \\
\hat{y}^{ATC} &= \hat{p}^{ctr} \times \hat{p}^{avr} \\
                &= p(y^{Click}=1|\mathbf{x}) \times p(y^{ATC}=1|y^{Click}=1, \mathbf{x}) \\
\hat{y}^{Purchase} &= \hat{p}^{ctr} \times \hat{p}^{avr} \times \hat{p}^{cvr} \\
                    &= p(y^{Click}=1|\mathbf{x}) \times p(y^{ATC}=1|y^{Click}=1, \mathbf{x}) \\
                    &\times p(y^{Purchase}=1|y^{Click}=1, y^{ATC}=1, \mathbf{x}) \\
\end{aligned}
\end{equation}
where $y^{Click}=1$, $y^{ATC}=1$, $y^{Purchase}=1$ denote whether click or add-to-cart or purchase event occurs, respectively. $\hat{y}^{Click}$, $\hat{y}^{ATC}$, and $\hat{y}^{Purchase}$ denote the final outputs of the model, respectively. $\hat{p}^{ctr} = p(y^{Click}=1|\mathbf{x})$ denotes the post-view click-through rate. $\hat{p}^{avr} = p(y^{ATC}=1|y^{Click}=1, \mathbf{x})$ denotes the click-through add-to-cart conversion rate, which is defined as the conditional probability of the product being added to cart given that it has been clicked. Similarly, $\hat{p}^{cvr} = p(y^{Purchase}=1|y^{Click}=1, y^{ATC}=1, \mathbf{x})$ denotes the click-through conversion rate, defined as the conditional probability of the product being purchased given that it has been added into cart, which depicts the complete behavior sequence: impression $\rightarrow$ click $\rightarrow$ add-to-cart $\rightarrow$ purchase.

\subsection{Loss Optimization}

The final loss is a linear combination of the losses of the individual tasks:

\begin{equation}
    \mathcal{L}_{MTL} = \sum_{k} w_k \cdot \mathcal{L}_k
\end{equation}
where $w_k$ is the task-specific weight and $\mathcal{L}_k$ is the task-specific loss function. In MLPR, we adopt the uncertainty weighting of the loss optimization \cite{b8} which uses the homoscedastic uncertainty to balance the single-task losses. The model's homoscedastic uncertainty or task-dependent uncertainty is not output but a quantity that remains constant for different input examples of the same task. The optimization procedure is carried out to maximize a Gaussian likelihood objective that accounts for the homoscedastic uncertainty. In the model the uncertainty loss can be formulated as:

\begin{equation}
\begin{aligned}
    \mathcal{L}_{MTL}(\mathbf{W},\sigma_1,\sigma_2,\sigma_3)=&\dfrac{1}{2\sigma_1^{2}}\mathcal{L}_1(\mathbf{W})+\dfrac{1}{2\sigma_2^{2}}\mathcal{L}_2(\mathbf{W})+\dfrac{1}{2\sigma_3^{2}}\mathcal{L}_3(\mathbf{W})\\
    &+\log\sigma_1\sigma_2\sigma_3\\
\end{aligned}
\end{equation}
where $\mathcal{L}_1$, $\mathcal{L}_2$, and $\mathcal{L}_3$ represent the losses of the three tasks, respectively. $\sigma_1$, $\sigma_2$, and $\sigma_3$ are the corresponding noise parameters and can balance the task-specific losses. The trainable parameters should be automatically updated during the training process.

\section{Experimental Setup}

\subsection{Dataset}

The e-commerce dataset was collected from Walmart.com during one continuous month in Oct 2020, which contains the user search queries, the corresponding products in the search results, and the user engagement data for each query-item pair including the number of clicks, the number of adding to the shopping cart, and the number of purchases. We filtered out the query-item pairs with less than or equal to five impressions. Then, we  divided the data into training, validation, and test sets, with the percentage of 80\%, 10\%, and 10\%, respectively. Each query-item pair is associated with one or more types of engagement: clicks, ATC, and purchases. Table~\ref{overall_dataset} shows the data statistics. 

\begin{table}[!htbp]
\begin{tabular}{cccc}
\toprule
\bf Query & \bf Items & \bf Query-Item pairs & \bf Impressions  \\
\hline
467,622 & 4,286,211 & 14,856,350 & 312,926,929 \\
\bottomrule
\end{tabular}
\caption{Statistics of the e-commerce dataset.}
\label{overall_dataset}
\end{table}

\subsection{Evaluation Metrics}

We aim to evaluate two aspects of the proposed work: prediction and  ranking. First of all, the proposed multi-task learning model predicts the probability for each query-item pair on each of the three types of user engagement (clicks, ATC, and purchase). We use Area Under the Curve (AUC) of Receiver Operating Characteristic (ROC) for the prediction tasks as it is widely used for evaluating classification/prediction models \cite{b50}. To evaluate the ranking results for each test query, we apply Normalized Discounted Cumulative Gain (NDCG) which is suitable for product search where users are usually sensitive to the ranked position of the relevant products\cite{b50}.

\begin{table*}[!htbp] 
\centering
\begin{tabular}{cccccccccc}
\toprule
\multicolumn{1}{c}{\multirow{2}*{\bf Model}}& \multicolumn{3}{c}{\bf AUC}& \multicolumn{3}{c}{\bf NDCG@1} & \multicolumn{3}{c}{\bf NDCG@5}\\
\multicolumn{1}{c}{}&\bf Click&\bf ATC&\bf Purchase &\bf Click&\bf ATC&\bf Purchase &\bf Click&\bf ATC&\bf Purchase\\
\hline
$MLP_{Single}$ & +3.93\% & +2.06\% & +0.01\% & +8.06\% & +3.39\% & +0.06\% & +5.46\% & +2.20\% & +1.36\% \\
$MLP_{MTL}$ & +3.78\% & +2.70\% & +0.03\% & +8.81\% & +4.97\% & -0.28\% & +5.85\% & +3.54\% & +1.36\% \\
$ESM^2$ & +1.48\% & +0.28\% & -0.70\% & +2.64\% & -2.66\% & -2.04\% & +1.42\% & -2.79\% & +0.03\% \\
MMoE & -0.73\% & -0.15\% & -1.01\% & -1.70\% & -5.23\% & -4.52\% & -1.66\% & -3.79\% & -1.99\% \\
PLE & +5.80\% & +3.63\% & +0.56\% & +10.14\% & +6.31\% & +3.28\% & +7.84\% & +4.86\% & +3.69\% \\
AITM & +5.86\% & +3.98\% & +0.64\% & +9.88\% & +6.93\% & +3.13\% & +7.73\% & +4.86\% & +3.49\% \\
\bf{MLPR} & $\bf+6.48\%^{\dagger}$ & $\bf+4.66\%^{\dagger}$ & $\bf+1.03\%^{\dagger}$ & $\bf+17.22\%^{\dagger}$ & $\bf+10.61\%^{\dagger}$ & $\bf+5.36\%^{\dagger}$ & $\bf+10.48\%^{\dagger}$ & $\bf+8.10\%^{\dagger}$ & $\bf+5.65\%^{\dagger}$ \\
\bottomrule
\end{tabular}
\caption{\bf Experimental results in terms of percentage lift  over XGBoost in AUC, NDCG@1, and NDCG@5 for the tasks: Click, Add-to-cart (ATC) and Purchase. The best results on each task are highlighted. $\dagger$ denotes statistically significant improvement from XGBoost to MLPR with the p-value $< 0.0001$ using the two-tailed t-test.}
\label{experimental_results}
\end{table*}

\subsection{Baseline Methods}
We compare MLPR with the following competitive baselines:

\begin{itemize}
\item \textbf{XGBoost} \cite{b49}: XGBoost is a gradient boosting framework that uses tree-based learning algorithms. It has been widely-used in industrial ranking systems. In the experiments, only relative metric improvements over XGBoost  instead of absolute values are presented, due to the company confidential policy.
\item $\mathbf{MLP_{Single}}$: This is a single-task learning model with the basic multi-layer perceptron (MLP) used for each task.
\item $\mathbf{MLP_{MTL}}$: We use shared-bottom structure at the bottom and tower network at the top. The structure of shared-bottom and tower network are multi-layer perceptron \cite{b46}.
\item $\mathbf{ESM^2}$ \cite{b6}: The ESMM \cite{b2} and $ESM^2$ with probability transfer pattern were designed for solving the non-end-to-end post-click conversion rate via training on the entire space to relieve the sample selection bias problem.
\item \textbf{MMoE} \cite{b4}: The MMoE with Expert-Bottom pattern is designed to integrate experts via multiple gates in the Gate Control.
\item \textbf{PLE} \cite{b5}: The Progressive Layered Extraction (PLE) with Expert-Bottom pattern separates task-shared experts and task-specific experts explicitly under different task correlations.
\item \textbf{AITM} \cite{b1}: The AITM model with adaptive information module transfers the knowledge from different conversion stages in the vector space. 
\end{itemize}

\section{Experimental Results}

\subsection{Baseline Comparison}

In this section, we compare the proposed MLPR model with the baseline methods on the three types of user engagement: clicks, add-to-cart (ATC), and purchases. Table~\ref{experimental_results} contains the results and we have the following observations. First of all, MLPR achieved the best results on all the three tasks in both metrics and outperformed all the competitive baselines with a significant margin. The deep learning based models yielded better results than the traditional XGBoost method, indicating the advantages of the neural methods in utilizing a large amount of data for model training. We can then compare a basic multi-task learning model $MLP_{MTL}$ with the single-task learning models $MLP_{Single}$ and XGBoost. The results showed that $MLP_{MTL}$ achieved better results in most of the metrics, which indicated the effectiveness of a multi-task learning approach in transferring knowledge between different tasks.

Based on the experts-bottom-based structure, the standard MMoE model could not perform well on the dataset. It performed even worse than the $MLP_{MTL}$ model, as it only controls the shared knowledge among different tasks. However, the PLE model with the specific-experts layer could improve significantly by transferring the shared information and task-specific knowledge among various tasks. The AUC of Click was improved by 5.8\% over XGBoost. The ${ESM^2}$ and AITM models optimize the performance in the upper level of the model structure. The simple probability transfer learning structure in ${ESM^2}$ transfers the knowledge with a simple conditional probability between adjacent tasks. The AITM model with attention module could obtain more gains by the sequential dependence tasks, which achieved competitive results. Our proposed MLPR model obtained significant improvement compared to various state-of-the-art baseline models and demonstrated the effectiveness of the proposed multi-task learning architecture with the domain-specific BERT for product ranking.

\subsection{Ablation Study}

This section discusses the effect of different components and stages of the MLPR model. 

\subsubsection{Domain-specific BERT with Fine-tuning}

We fine-tune the domain-specific BERT with the downstream multi-task learning. The experimental results from Table~\ref{fine_tuning} demonstrate that the fine-tuning of BERT has significant improvement on each prediction task, either with the basic $MLP_{MTL}$ model or with the MLPR model. Especially on the CTR prediction task, even the basic $MLP_{MTL}$ model with fine-tuning yielded $1.41\%$ of improvement over the one without fine-tuning in AUC score, and $7.03\%$ of improvement in NDCG@1.

We investigated some specific queries and their search results from XGBoost and our model respectively. For example, given the query ``half bed for kids", the top result returned by XGBoost was the product with the title ``The cincinnati Kid POSTER (22x28) (1965) (Half Sheet Style A)", due to the lexical matching between the query and the title, but this item was not relevant at all. On the other hand, our model returned the product ``Bedz King Stairway Bunk Beds Twin over Full with 4 Drawers in the Steps and a Twin Trundle, Gray" as the top result. As we can see, the title did not have as much lexical overlap with the given query as the previous product had, but it was semantically relevant to the query. This example demonstrated the effectiveness of the proposed BERT based method in bridging the vocabulary gap.

\begin{table*}[!htbp] 
\centering
\begin{tabular}{cccccccccc}
\toprule
\multicolumn{1}{c}{\multirow{2}*{\bf Model}}& \multicolumn{3}{c}{\bf AUC} & \multirow{2}*{\bf AUC Gain} & \multicolumn{3}{c}{\bf NDCG@1} & \multirow{2}*{\bf NDCG@1 Gain}\\
\multicolumn{1}{c}{}&\bf Click&\bf ATC&\bf Purchase&{}&\bf Click&\bf ATC&\bf Purchase&{}& \\
\hline
$MLP_{MTL}$ W/O Fine-tuning & +3.77\% & +2.78\% & +0.20\% &  N/A, N/A, N/A & +8.28\% & +4.68\% & +0.28\% &  N/A, N/A, N/A \\
$MLP_{MTL}$ With Fine-tuning & +5.23\% & +3.24\% & +0.59\% & 1.41\%,0.45\%,0.39\% & +15.91\% & +9.03\% & +3.93\% & 7.03\%, 4.14\%, 3.74\% \\
\hline
MLPR W/O Fine-tuning & +5.63\% & +4.05\% & +0.66\% &  N/A, N/A, N/A & +10.14\% & +7.22\% & +3.28\% &  N/A, N/A, N/A \\
{MLPR} With Fine-tuning & +6.48\% & +4.66\% & +1.03\% &0.81\%,0.58\%,0.37\%& +17.22\% & +10.61\% & +5.36\% & 6.44\%,3.19\%,2.02\% \\
\bottomrule
\end{tabular}
\caption{\bf Experimental results of the domain-specific BERT with fine-tuning vs without (W/O) fine-tuning in the basic MTL model and the MLPR model. AUC and NDCG@1 are reported in terms of the percentage lift over XGBoost. AUC Gain and NDCG@1 Gain are reported in terms of the percentage lift for fine-tuning over without fine-tuning.}
\label{fine_tuning}
\end{table*}

\begin{table*}[!htbp] 
\centering
\begin{tabular}{ccccccccc}
\toprule
\multirow{2}*{\bf Model} & \multicolumn{3}{c}{\bf AUC} & \multirow{2}*{\bf AUC Gain} & \multicolumn{3}{c}{\bf NDCG@1} & \multirow{2}*{\bf NDCG@1 Gain}\\
{} & \bf Click&\bf ATC &\bf Purchase & {} &\bf Click&\bf ATC &\bf Purchase & {} \\
\hline
$MLP_{MTL}$ & +3.78\% & +2.70\% & +0.03\% & N/A, N/A, N/A & +8.81\% & +4.97\% & -0.28\% &  N/A, N/A, N/A \\
+Uncertainty Loss & +3.77\% & +2.78\% & +0.20\% & -0.01\%, 0.08\%, 0.17\% & +8.28\% & +4.68\% & +0.28\% & -0.48\%, -0.28\%, 0.56\% \\
+Specific-Experts & +5.80\% & +3.63\% & +0.56\% &1.94\%, 0.91\%, 0.53\% & +10.14\% & +6.31\% & +3.28\% &1.22\%, 1.26\%, 3.57\%\\
+Attention Units & +5.39\% & +3.78\% & +0.63\% &1.55\%, 1.06\%, 0.60\%& +9.85\% & +6.80\% & +3.31\% & 0.95\%, 1.75\%, 3.60\%\\
+Probability Transfer & +5.69\% & +4.03\% & +0.67\% &1.84\%, 1.30\%, 0.64\%& +9.85\% & +6.85\% & +3.31\% &0.96\%, 1.80\%, 3.59\%\\
+Fine-tuning & \bf+6.48\% & \bf+4.66\% & \bf+1.03\% &\bf2.60\%, \bf1.91\%, \bf1.00\%& \bf+17.22\% & \bf+10.61\% & \bf+5.36\% & \bf7.74\%, \bf5.39\%, \bf5.65\%\\
\bottomrule
\end{tabular}
\caption{\bf Experimental results of incrementally adding individual components to the base $MLP_{MTL}$ model. AUC and NDCG@1 are reported in terms of the percentage lift over XGBoost. AUC Gain and NDCG@1 Gain are reported in terms of the percentage lift over the base $MLP_{MTL}$ model. The best results on each task are highlighted.}
\label{module_combination}
\end{table*}

\subsubsection{The Effect of Different Stages of MLPR}

In order to understand the performance of each stage of our framework, we investigate the individual components of the MLPR model by incrementally adding a new component to the base multi-task learning model. We define each  component in our model as follows:

\begin{itemize}
    \item $\mathbf{MLP_{MTL}}$ The base model. The shared-bottom stage design as multi-layer feed-forward network and the upper stage design as tower network for each task.
    \item \textbf{+Uncertainty Loss} The same structure with $MLP_{MTL}$ but with a different loss function (uncertainty loss) for training.
    \item \textbf{+Specific-Experts} Based on the previous $MLP_{MTL}$ structure and the uncertainty loss, the model adds a new component (specific-experts with customized gating) to the shared-bottom stage. 
    \item \textbf{+Attention Units} Based on the previous model structure, the model adds the attention units after the tower model in the upper stage of the model.
    \item \textbf{+Probability Transfer} The model implements the probability transfer component based on the previous design, which regularizes the predicted results from the attention units.
    \item \textbf{+Fine-tuning} The fine-tuning process is applied based on the previous model. The parameter within the domain-specific BERT will be updated by the top-level optimization function.
\end{itemize}

Table~\ref{module_combination} contains the experimental results. As we can see, with the Uncertainty Loss, the model obtained slightly better while comparable results overall. Further with the Specific-Experts layer, the model performance significantly improved, especially for the AUC score of CTR prediction, with a 1.94\% increase. Because the specific experts could extract more confidential information than the simple shared-bottom design, the specific experts stage not only extracts the common knowledge from different tasks but also learns the specific information for each individual task. The model with Attention Units on the upper level also demonstrated good improvement over the base model. Moreover, the probability transfer component showed positive results, as it optimized the joint predictions for the multiple tasks. Last but not the least, with the benefits of domain-specific BERT, the model learned valuable information from the text field. With the fine-tuning process, the model gained the best result, which demonstrated the effectiveness of using the fine-tuned BERT for product search.

\begin{table}[t]
\begin{tabular}{ccc}
\toprule
\bf Model & \bf Deployment Strategy & \bf Time \\
\hline
XGBoost & W/O query/product embedding & 58ms \\
$MLP_{MTL}$ & product pre-computing & 96ms \\
MLPR & W/O product pre-computing & 171ms \\
MLPR & product pre-computing & 112ms \\
\bottomrule
\end{tabular}
\caption{\bf The latency in milliseconds (ms) at 99 percentile on product ranking.}
\label{latency_performance}
\end{table}

\subsection{Latency Performance}

To understand the efficiency of MLPR in inference, we conducted an analysis on latency by experimenting with four models. We used the 99th-percentile (P99) of product ranking time as the latency metric, which was measured from the time the model received the query to the time it returned a ranked list of 100 products. The experiments were performed on an Intel(R) Xeon(R) CPU E5-2660 v4 @ 2.00GHz machine and NVIDIA Tesla V100 GPU with 16G memory.

The offline P99 latency on the test set is reported in Table~\ref{latency_performance}. As we can see, XGBoost had the lowest running time among the four models, since it did not compute query and product embeddings. The MLPR model could save a significant amount of time when using pre-computing of product embeddings, as the inference time was dropped from 171ms to 112ms. With product pre-computing, MLPR was slightly slower than $MLP_{MTL}$ while MLPR has a more sophisticated architecture with a higher accuracy as shown in Table \ref{experimental_results}. The experimental results demonstrated the efficiency of the proposed multi-task learning with the pre-computing strategy.

\subsection{Transfer Knowledge Gain}

In order to understand the transfer knowledge gained through different tasks, we compared the model performance with different sampling strategies to demonstrate that our model has a robust generalization ability. Firstly, we sorted the query-item pairs in the test dataset according to the number of impressions and divided the test dataset into three groups based on the percentiles, i.e., top 0\%-25\%, 25\%-75\%, and 75\%-100\% of the sorted test dataset. The top 0\%-25\% data corresponds to the portion of the test query-item pairs with the least impressions and the top 75\%-100\% data includes the test instances with the most impressions. 

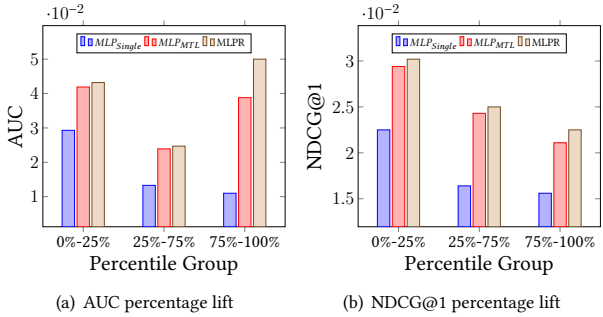
\begin{figure}[t]
\subfigure[AUC percentage lift]{
  \begin{tikzpicture}[scale=0.47]
\begin{axis}
[
    ybar, 
    enlargelimits=0.25,
    legend style={at={(0.5, 0.96)}, 
      anchor=north,legend columns=-1},   
    ylabel={AUC}, 
    xlabel={Percentile Group},
    label style={font=\Huge},
    tick label style={font=\huge},
    symbolic x coords={0\%-25\%, 25\%-75\%, 75\%-100\%},
    xtick=data,
    ]
\addplot coordinates {(0\%-25\%, 0.0293 ) (25\%-75\%, 0.0133 ) (75\%-100\%, 0.0110 )}; 
\addplot coordinates {(0\%-25\%, 0.0419 ) (25\%-75\%, 0.0239 ) (75\%-100\%, 0.0388 )};
\addplot coordinates {(0\%-25\%, 0.0432 ) (25\%-75\%, 0.0247 ) (75\%-100\%, 0.0500 )};
\legend{$MLP_{Single}$, $MLP_{MTL}$, MLPR}
\end{axis}
\end{tikzpicture}
  \label{fig:circle} 
}
\subfigure[NDCG@1 percentage lift]{
  \begin{tikzpicture}[scale=0.47]
\begin{axis}
[
    ybar, 
    enlargelimits=0.25,
    legend style={at={(0.5, 0.96)}, 
      anchor=north,legend columns=-1},   
    ylabel={NDCG@1}, 
    xlabel={Percentile Group},
    symbolic x coords={0\%-25\%, 25\%-75\%, 75\%-100\%},
    label style={font=\Huge},
    tick label style={font=\huge},
    xtick=data,
    ]
\addplot coordinates {(0\%-25\%, 0.0225 ) (25\%-75\%, 0.0164 ) (75\%-100\%, 0.0156 )};
\addplot coordinates {(0\%-25\%, 0.0294 ) (25\%-75\%, 0.0243 ) (75\%-100\%, 0.0211 )};
\addplot coordinates {(0\%-25\%, 0.0302 ) (25\%-75\%, 0.0250 ) (75\%-100\%, 0.0225 )};
\legend{$MLP_{Single}$, $MLP_{MTL}$, MLPR}
\end{axis}
\end{tikzpicture}
  \label{fig:rectangle} 
}
\caption{Transfer knowledge gain of each percentile group with different models over XGBoost model.}
\label{knowledge_gain}
\end{figure}

As the results shown in Figure~\ref{knowledge_gain}, our proposed model demonstrated relatively stable performance across the three different groups. It could achieve reasonable performance even when less engagement data was used. On the other hand, compared to the baseline models, our model has improved on different tasks in both AUC and NDCG metrics, especially in the top 0\%-25\% group. 

\subsection{Hyperparameter Analysis}

In this section, we perform an analysis on two important hyper-parameters of the proposed deep learning architecture: the dropout ratio and the number of the hidden layers.

\subsubsection{Dropout Ratio}

If the model has too many parameters and too few training samples, the trained model is likely to overfit \cite{b9}. As a widely used technique to alleviate overfitting in neural nets, the dropout mechanism \cite{JMLR:v15:srivastava14a} can randomly deactivate some neural nodes. This method can diminish the interaction between hidden layer nodes, and improve the model's generalization ability. We experimented with different dropout ratios, ranged from 0.2 to 0.8. As we can see from the results shown in Figure~\ref{drop_ratio}, when the dropout ratio is 0.2, the model obtained the best performance. In all the other experiments, we used 0.2 as the dropout ratio for MLPR unless specified otherwise.

\subsubsection{Number of the Hidden Layers and Nodes}

In a deep neural model, increasing the number of layers of the network can generally increase the model capacity \cite{b10}. However, it will also increase the number of model parameters, which may result in overfitting. In the experiments, we tried different number of layers ranged from 2 to 4 in the MLP component of the underlying expertise network. As shown in Figure~\ref{hidden_layers}, with the increase of the number of layers, the model performance was improved at the beginning but then declined, which indicated the network may start to overfit. Thus, we chose a 3-layer network as the MLP component structure. In addition, we tested the number of nodes in different hidden layers. We found that with the increase of the number of hidden layer nodes, the model generally performed better. In all the other experiments, we used [512, 256, 128] as the numbers of the nodes in the hidden layers.

\pgfplotsset{
    compat=1.16, 
    every axis/.append style={scale only axis, axis on top,
    height=4.25cm, width=3cm, xmin=-1, xmax=240,
    }
}

\begin{figure}
\subfigure[Dropout ratio]{
\begin{tikzpicture}[scale=.95]
\begin{axis}[
    name=plot1,
    legend style={nodes={scale=0.8, transform shape}},
    legend style={at={(0.35,0.45)},anchor=north},
    ylabel={AUC},
    xmin=0.2, xmax=0.8,
    ymin=0, ymax=0.06,
    xtick={0.2,0.3,0.4,0.5,0.6,0.7,0.8},
    ytick={0,0.01,0.02,0.03,0.04,0.05,0.06},
    ymajorgrids=true,
    grid style=dashed,
]

\addplot[
    color=blue,
    mark=square,
    ]
    coordinates {
    (0.2,0.0555 )(0.3,0.0546 )(0.4,0.0523 )(0.5,0.0497 )(0.6,0.0476 )(0.7,0.0436 )(0.8,0.0389 )
    };
\addplot[
    color=red,
    mark=*,
    ]
    coordinates {
    (0.2,0.0389 )(0.3,0.0382 )(0.4,0.0362 )(0.5,0.0344 )(0.6,0.0326 )(0.7,0.0293 )(0.8,0.0261 )
    };
\addplot[
    color=black,
    mark=triangle,
    ]
    coordinates {
    (0.2,0.0066 )(0.3,0.0063 )(0.4,0.0060 )(0.5,0.0056 )(0.6,0.0050 )(0.7,0.0041 )(0.8,0.0022 )
    };
    \legend{Click, ATC, Purchase}
    
\end{axis}
\end{tikzpicture}
\label{drop_ratio}
}
\subfigure[The number of hidden layers]{
\begin{tikzpicture}[scale=.95]
\begin{axis}[%
    legend style={nodes={scale=0.8, transform shape}},
    legend style={at={(0.35,0.45)},anchor=north},
    name=plot2,
    xmin=2, xmax=4,
    ymin=0, ymax=0.06,
    xtick={2,3,4},
    ytick={0,0.01,0.02,0.03,0.04,0.05,0.06},
    ymajorgrids=true,
    grid style=dashed,
]
\addplot[
    color=blue,
    mark=square,
    ]
    coordinates {
    (2, 0.0555 )(3,0.0563 )(4,0.0539 )
    };
\addplot[
    color=red,
    mark=*,
    ]
    coordinates {
    (2, 0.0389 )(3,0.0405 )(4,0.0379 )
    };
\addplot[
    color=black,
    mark=triangle,
    ]
    coordinates {
    (2, 0.0066 )(3,0.0066 )(4,0.0061 )
    };
    \legend {Click, ATC, Purchase}
\end{axis}
\end{tikzpicture}
\label{hidden_layers}
}
\caption{The AUC percentage lift of the MLPR model over the XGBoost model with various values of the dropout ratio and the number of hidden layers.}
\end{figure}
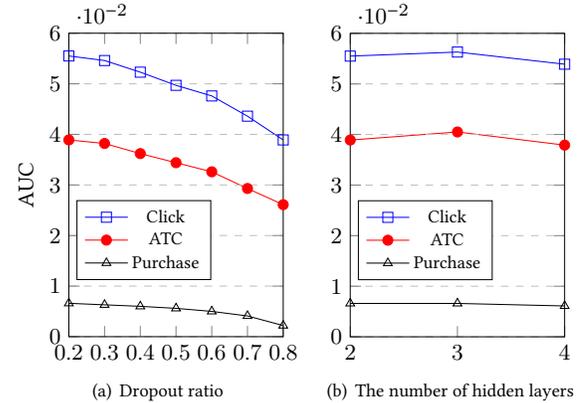

\section{Conclusion and Future Work}

In this paper, we propose a novel multi-task learning framework for product ranking by integrating multiple types of engagement signals with neural information retrieval. The proposed end-to-end learning framework incorporates a domain-specific BERT for semantic match with traditional ranking features. The comprehensive experiments on a real-world e-commerce dataset have demonstrated the effectiveness of the proposed approach.

This work is an initial step towards a promising research direction. The proposed framework allows flexible configuration, such as input data, text embedding extraction, mixture-of-experts, loss function, etc. In the future, we will integrate other types of input data into the model such as product images. In addition, we will conduct A/B testing to validate the performance of the proposed model in the online environment. Last but no the least, we will apply the proposed framework to other search and ranking tasks.

\bibliographystyle{ACM-Reference-Format}
\bibliography{main}

\end{document}